\documentclass{astlb}
\usepackage[utf8]{inputenc}
\usepackage[english]{babel}
\usepackage{graphicx}
\graphicspath{ {images/} }
\usepackage{float} 
\usepackage{wrapfig}
\usepackage[dvipsnames]{xcolor}
\usepackage{bm}
\usepackage{amsmath}
\usepackage{caption, subcaption}
\usepackage{booktabs}
\usepackage{siunitx}
\usepackage{ulem}
\usepackage{amssymb}
\usepackage{tikz}
\usepackage{mathtools, cuted}
\usepackage{calrsfs}
\usepackage{natbib}
\usepackage{xspace}
\usetikzlibrary{trees}
\usetikzlibrary{arrows.meta, chains, positioning, quotes}
\DeclareMathAlphabet{\pazocal}{OMS}{zplm}{m}{n}
\usepackage[hyperfootnotes=false]{hyperref}

\usepackage{float}
\usepackage{multicol}
\usepackage{url}

\usepackage{enumitem}

\newcommand{\srge}{{SRG/eROSITA}\xspace}
\newcommand{\ergcms}{erg/s/cm$^2$\xspace}

\newcommand{\ntotal}{{6885}\xspace}
\newcommand{\nhostless}{{539}\xspace}
\newcommand{\noverpany}{{6346}\xspace}

\newcommand{\nunknownnozmag}{{60}\xspace}

\newcommand{\nextragal}{{5929}\xspace}
\newcommand{\ngal}{{357}\xspace}
\newcommand{\nunknown}{{3462}\xspace}
\newcommand{\nwithclasses}{{2884}\xspace}
\newcommand{\nvalidation}{{2864}\xspace}

\begin{document}

\journalinfo{2022}{48}{12}{755}[766]
\title{SRG/eROSITA Survey in the Lockman Hole: Classification of X-ray Sources}

\author{
M.I. Belvedersky\address{1,2} \email{belveder@cosmos.ru},
S.D. Bykov\address{3, 4},
M.R. Gilfanov\address{1,4},
  \addresstext{1}{Space Research Institute, Russian Academy of Sciences, Profsoyuznaya 84/32, 117997 Moscow, Russia}
  \addresstext{2}{National Research University Higher School of Economics (HSE), Moscow, Russia}
  \addresstext{3}{Kazan Federal University, Department of Astronomy and Satellite Geodesy, 420008 Kazan, Russia}
  \addresstext{4}{Max Planck Institute for Astrophysics, Karl-Schwarzschild-Str 1, Garching b. Muenchen D-85741, Germany}
}

\shortauthor{Belvedersky, Bykov, \& Gilfanov}
\shorttitle{Classification of SRG/eROSITA sources in Lockman Hole survey}
\submitted{15.11.2022}

\begin{abstract}
We have classified the point-like X-ray sources detected by the SRG/eROSITA telescope in the deep Lockman Hole survey. The goal was to separate the sources into Galactic and extragalactic objects. In this work have used the results of our previous cross-match of X-ray sources with optical catalogs. To classify SRG/eROSITA sources we have used the flux ratio $F_{\rm x}/F_{\rm o}$ and information about the source optical extent. As a result, of the 6885 X-ray sources in the eROSITA catalog 357 sources have been classified as Galactic and 5929 and as extragalactic. 539 out of 6885 have been treated as hostless, i.e., having no optical counterparts in the optical catalogs under consideration. 60 have remained unclassified due to the insufficient reliability of optical photometry. Recall and precision for the extragalactic sources are 99.9 and 98.9\% (respectively) and 91.6 and 99.7\% for the Galactic sources. Using this classification, we have constructed the curves of cumulative number counts for the Galactic and extragalactic sources in the Lockman Hole field. The code that accompanies this paper is available at \url{https://github.com/mbelveder/ero-lh-class.git}.


\end{abstract}

\section{INTRODUCTION}
The Spectrum-RG (SRG) orbital X-ray observatory was launched on July 13, 2019, and is currently in a halo orbit around the libration point L2 of the Sun–Earth system \citep{sunyaev2021}. There are two X-ray telescopes with grazing-incidence optics onboard the observatory: the eROSITA telescope \citep{predehl2021} sensitive in the 0.2–9.0 keV energy band and the Mikhail Pavlinsky ART-XC telescope \citep{pavlinsky2021} sensitive in the 4–30 keV energy band. It is expected that during its 4-year all-sky survey SRG/eROSITA will detect $\sim4$ million X-ray sources of various types in the entire sky and will provide a huge volume of data to solve a wide range of problems in astrophysics and cosmology \citep{Prokopenko2009, Merloni2012, Kolodzig2013a, Kolodzig2013b}. ART-XC investigates hard X-ray objects \citep{Mereminskiy2018, Mereminskiy2019}.

To study the populations of objects of various types detected in X-ray surveys, first of all it is necessary to separate the objects of our Galaxy (stars, compact objects) and extragalactic objects (galaxies, active galactic nuclei -- AGNs, quasars). The problem is complicated by the fact that the data on the distances to X-ray objects are often unavailable \citep{Salvato2019}. Therefore, we to use additional multiwavelength information (for example, optical spectra or photometry) to determine whether a particular source is Galactic or extragalactic. Some existing methods include the diagnostics based on the spectral energy distribution (SED) \citep{Salvato2009, Marchesi2016}, the classification by positions on the color–magnitude or color–color diagrams \citep{Maccacaro1988, Brusa2007, Brusa2010, Xue2011, Civano2012, Salvato2018, Salvato2022}, and the identification of sources with proper motions based on data from astrometric missions such as Gaia \citep{GaiaCollaboration2016, GaiaCollaboration2022}. For example, a multilevel scheme in which the main factors of the classification into Galactic and extragalactic objects are the object extent in an optical image, its proper motion, and the relation between the optical ($grz$), infrared ($W1$), and X-ray (0.5–2 keV) fluxes is used in \cite{Salvato2022} to solve this problem. To evaluate the quality of the classification schemes, it is necessary that some of the objects have a known class through spectroscopy or other reliable methods.

The goal of this paper is to classify the point X-ray sources detected by eROSITA during the Lockman Hole survey. For this task, apart from X-ray data, we use data from the Gaia astrometric satellite and the DESI LIS photometric survey. Our method is distinguished by the simplicity of interpretation and can be used to construct more complex classification algorithms, including those for X-ray all-sky survey data.

The paper is structured as follows. In Section 2 we describe the data used. In Section 3 we present the sample of sources with reliable information about their distances and our algorithm for the classification of sources with unknown distances. We describe and discuss the results in Section 4 and reach the conclusions in Section 5. The magnitudes used are in the AB system.

\section{DATA}\label{sect-data}

\subsection{X-ray data}\label{sect-data-xray}

The SRG/eROSITA Lockman Hole survey was performed in October 2019 during the verification observations of the SRG observatory. The survey characteristics, the procedure for constructing the catalog of X-ray sources, and the catalog itself are presented in \cite{gilfanov2022}. The catalog of X-ray sources includes 6885 point-like sources with a detection likelihood (\texttt{DET\_LIKE}) $>$ 10, roughly corresponding to a 4$\sigma$ significance.

\subsection{Search for Optical Counterparts of X-ray Sources}
The most important step before the classification of X-ray sources is their cross-match with optical catalogs. This problem is not trivial, since the error circle of X-ray sources often include more than one optical object; the exact number depends on the depth of the optical catalog and the sky density of sources in the area. We chose the DESI Legacy Imaging Surveys DR9 catalog (\citealt{dey2019}; hereafter DESI LIS) as a reference optical catalog. The eROSITA X-ray sources were cross-matched with optical objects from the DESI LIS catalog in \citet{bykov}. For this purpose, we used a neural network model to characterize the photometric attributes of optical counterparts to X-ray sources and field objects (not counterparts) in combination with the NWAY code \citep{Salvato2018}. The Nway algorithms are based on the Bayesian formalism and allow any additional (apart from positional and photometric) prior information to be used to search for counterparts. In \citet{bykov} for this purpose we used predictions of the classifier trained on a specially prepared sample. As a result, we found the most probable optical DESI LIS counterpart for each point-like SRG/eROSITA source.

For each X-ray source the Lockman Hole catalog provides the parameter $p_{\rm \text(any)}$ that characterizes our confidence in the presence of an optical counterpart \citep{bykov}. The closer this parameter to unity, the more reliable the assertion that the X-ray source has a counterpart in the optical catalog under consideration. The objects with low $p_{\rm \text(any)}$ are likely \textit{hostless}, i.e., they have no counterpart in the optical catalog. Following the recommendation given in \citet{bykov}, we chose an upper boundary for the classification of the source as hostless, $p_{\rm \text(any,0)}$ = 0.12. With this choice 539 sources (7.8\%) are classified as hostless. We plan to study these objects in the future. Here we will investigate the sources classified as having a counterpart in the DESI LIS catalog, their total number is 6346. The classification precision and the choice of the $p_{\rm \text(any)}$ threshold are discussed in detail in \citealt{bykov}.

\subsection{Photometric, Astrometric, and Spectroscopic Data}

Photometric information about the counterparts was taken from the DESI LIS catalog. This catalog includes the surveys performed with three telescopes: BASS (in the $g$, $r$ bands), MzLS ($z$), and DECaLS ($g$, $r$, $z$). For the Lockman Hole we used the BASS and MzLS data, since the DECaLS survey was not conducted in this region. All of the fluxes were corrected for interstellar absorption.

Information about the proper motions of the objects was taken from the Gaia DR3 catalog (hereafter Gaia) \citealt{GaiaCollaboration2016, GaiaCollaboration2022}.

Spectroscopic information is available for $\sim$ 1/3 of the X-ray catalog. For them we used the spectral classification and redshifts from the Sloan Digital Sky Survey DR16 (\citealt{SDSS2020}, hereafter SDSS) and The Million Quasars Catalogue\footnote{\url{https://heasarc.gsfc.nasa.gov/W3Browse/all/milliquas.html}} v7.5 (\citealt{Flesch21}, hereafter Milliquas) catalogs. The SDSS DR16 catalog is the fourth release of the fourth SDSS phase (SDSS-IV). The Milliquas catalog is a composite one and contains information about quasars and AGNs from various sources published up until April 30, 2022. We additionally used the Simbad database \citep{SIMBAD2000}, which also contains the information collected from various sources. Note that the Simbad database is not a catalog and, therefore, the information from it will have a low priority for us compared to other sources of information.

\section{CLASSIFICATION OF X-RAY SOURCES}\label{sect-method}
\noindent
\subsection{Cross-Match with the Gaia Catalog and Spectrometric Catalogs}\label{sect-method-gaiaetc}

To classify X-ray sources we used the coordinates of their most probable optical counterparts from the DESI LIS catalog. For the classification of Galactic sources we used information about the source parallaxes and proper motions from the Gaia DR3 catalog. The optical counterparts were cross-matched with Gaia sources; the search was conducted within a \ang{;;0.5} radius. A counterpart in the Gaia catalog was found for 2475 DESI LIS objects. There were 2 Gaia sources in the search circle in 8 cases, the nearest one was chosen as a counterpart. The sources for which the signal-to-noise (S/N) exceeded 5 at least for one of the following parameters were classified as Galactic:

\begin{itemize}  
    \item \textit{pmra}, the proper motion in right ascension;
    \item \textit{pmdec}, the proper motion in declination;
    \item \textit{pm}, the total proper motion  ($\text{pm} = \sqrt{\text{pmra}^2 + \text{pmdec}^2}$);
    \item \textit{parallax}, the absolute stellar parallax.
\end{itemize}

We found a total of 338 Galactic sources. Following the recommendation of the DESI team, no correction for the proper motion of the Gaia sources was made due to the small (six months) difference between the reference epochs of the DESI and Gaia DR3 catalogs (D. Lang, the DESI team, personal communication).

Then we identified the sources for which there was an optical spectroscopy. For this purpose, we used the SDSS, Milliquas, and Simbad spectroscopic data. The search for matches was performed with a search radius of \ang{;;1}. In the case of SDSS and Milliquas, there were no more than one source in the search circle. For Simbad there were two and three sources in the search circle in 72 and 21 cases, respectively. At the same time, only four
objects with ambiguous matches in Simbad were not cross-matched with the Gaia, SDSS, or Milliquas catalogs. For these objects, as in the case of Gaia, we chose the nearest source. The final classification of objects was made with the following priority: Gaia, SDSS, Milliquas, and Simbad. This means that if the object under consideration was not classified as Galactic (according to the Gaia data) then its class was set in accordance with the SDSS data\footnote{Among the 144 galaxies found using the SDSS catalog, 7 had the AGN subclass. We designated these objects as quasars.} and so on. A priority higher than Milliquas was given to the SDSS catalog. Although Milliquas contains only reliable spectroscopic measurements, it was compiled from the catalogs produced by various authors (i.e., it is methodologically inhomogeneous). In our case, the result of the final classification changes insignificantly when a higher priority than SDSS is assigned to the Milliquas catalog. Thus, in total, we determined the classes for 2884 of the 6346 objects.

\subsection{$F_{\rm x}/F_{\rm opt}$ -- Color Diagram} \label{sect:fxfo_color}

To classify the remaining objects, we investigated the characteristics of the objects with a known class (see previous subsection). Of all the combinations of parameters, the logarithm of the ratio of the X-ray (0.5 -- 2 keV) and optical fluxes ($z$) in combination with the optical ($r$ -- $z$) color and the information from DESI LIS about the object optical extent turned out to be astrophysically most informative and interesting for our goal.

\begin{figure*}
\centering
\includegraphics[scale=0.55]{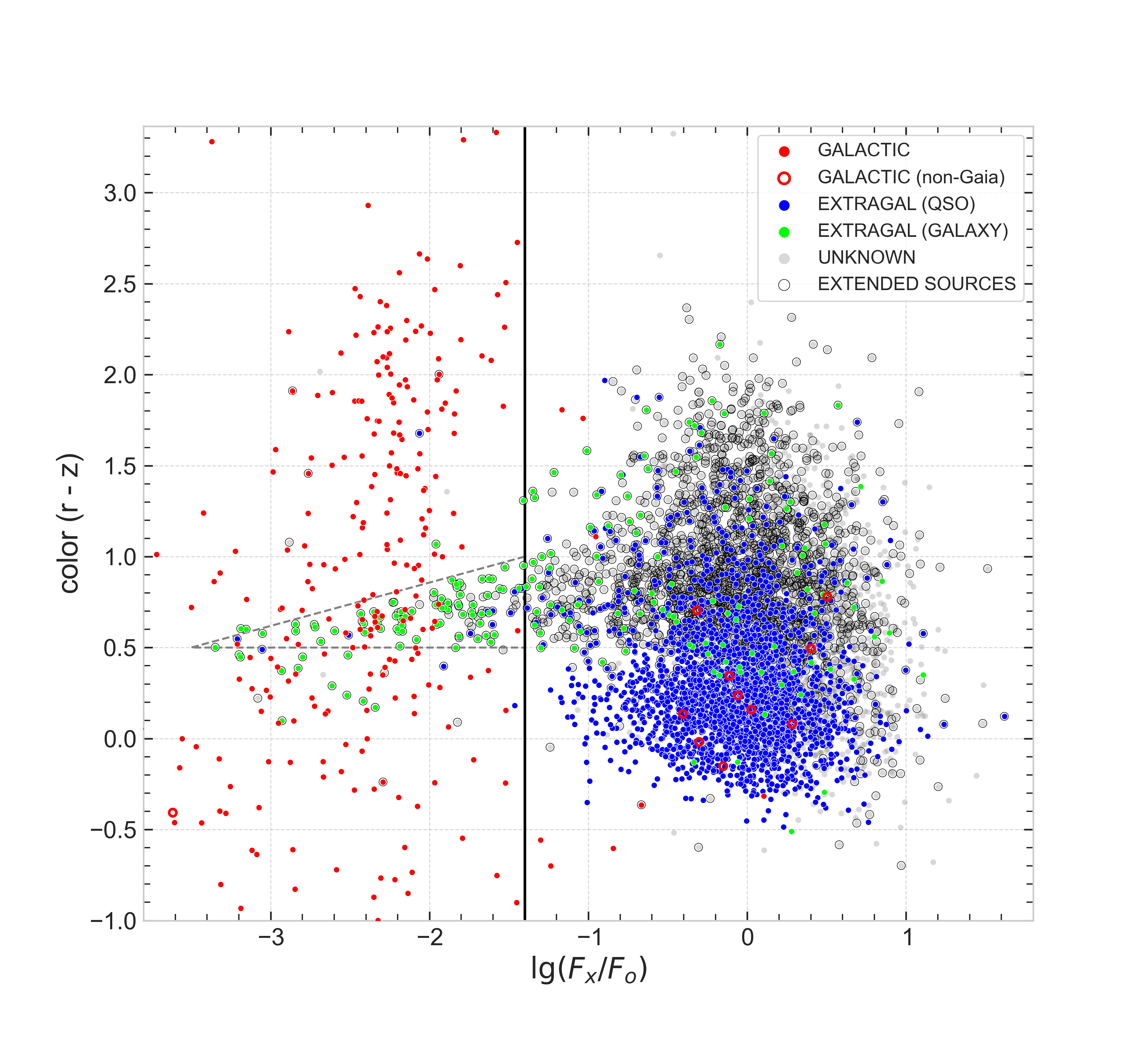}
\caption{
The $\lg(F_{\rm x} / F_{\rm o})$–optical color diagram for the SRG/eROSITA point-like X-ray sources in the Lockman Hole. The difference between the DESI LIS $r$ and $z$ magnitudes is plotted along the vertical axis. The logarithm of the ratio of the 0.5–2 keV X-ray flux to the optical DESI LIS $z$ flux is plotted along the horizontal axis. The objects classified based on the Gaia data or optical spectroscopy are indicated by different colors, according to the legend. The gray circles and the black circumferences designate the objects of unknown nature and the optically extended objects, respectively. Most of the Galactic objects are located in the left part of the diagram (leftward of the boundary marked by the solid vertical line). The extragalactic objects are located predominantly in the right part of the diagram, but some of them penetrate into the left part and are concentrated in the triangular region bounded by the dashed line.
}
\label{fig:FxFo_rz}
\end{figure*}

In Fig. \ref{fig:FxFo_rz} the X-ray sources are plotted on the $\lg(F_{\rm x} / F_{\rm o})$ -- optical color ($r$ -- $z$) plane. The objects whose nature was confirmed based on the Gaia measurements and those who have optical spectroscopy are designated by different colors. The sources that we failed to classify based on the spectroscopic data and the information about their proper motions are marked by the gray circles. The black circumferences mark the sources that are extended ones, according to the DESI LIS catalog ($\text{type} \neq \text{PSF}$).

The following conclusions can be drawn from an analysis of Fig. \ref{fig:FxFo_rz}.

\begin{enumerate}
    \item It can be seen that the Galactic objects (red circles) are located predominantly in the left part of this diagram, in the region of low $\lg(F_{\rm x} / F_{\rm o})$, while the extragalactic ones (green and blue circles) are located on the right. This is an expression of the well-known fact that the stars, on average, have lower ratios $\lg(F_{\rm x} / F_{\rm o})$ than the AGNs and quasars \citep{Maccacaro1988, Marchesi2016, Salvato2018, Hasinger2021}. Note that 57 Galactic sources turned out to be beyond the lower boundary in Fig. \ref{fig:FxFo_rz}. All these sources are brighter than 15 mag in the $g$ band and may be overexposed in the DESI LIS images, so that their color determined from the DESI LIS data may be distorted. The Gaia data unambiguously point to their Galactic nature, while the possible distortions of their optical color do not affect the classification of sources.
    
    Fig. \ref{fig:FxFo_valid_hist} presents the $\lg(F_{\rm x} / F_{\rm o})$ distributions for the Galactic (red solid line) and extragalactic (the green dash–dotted and blue dashed lines for the galaxies and quasars, respectively) sources. Based on Figs. \ref{fig:FxFo_rz} and \ref{fig:FxFo_valid_hist}, we introduced an arbitrary boundary, $\lg(F_{\rm x} / F_{\rm o})$ = -1.4, separating the stars and quasars. This boundary is indicated in Fig. \ref{fig:FxFo_rz} by the solid vertical line.

    \begin{figure*}
    \centering
    \includegraphics[scale=0.6]{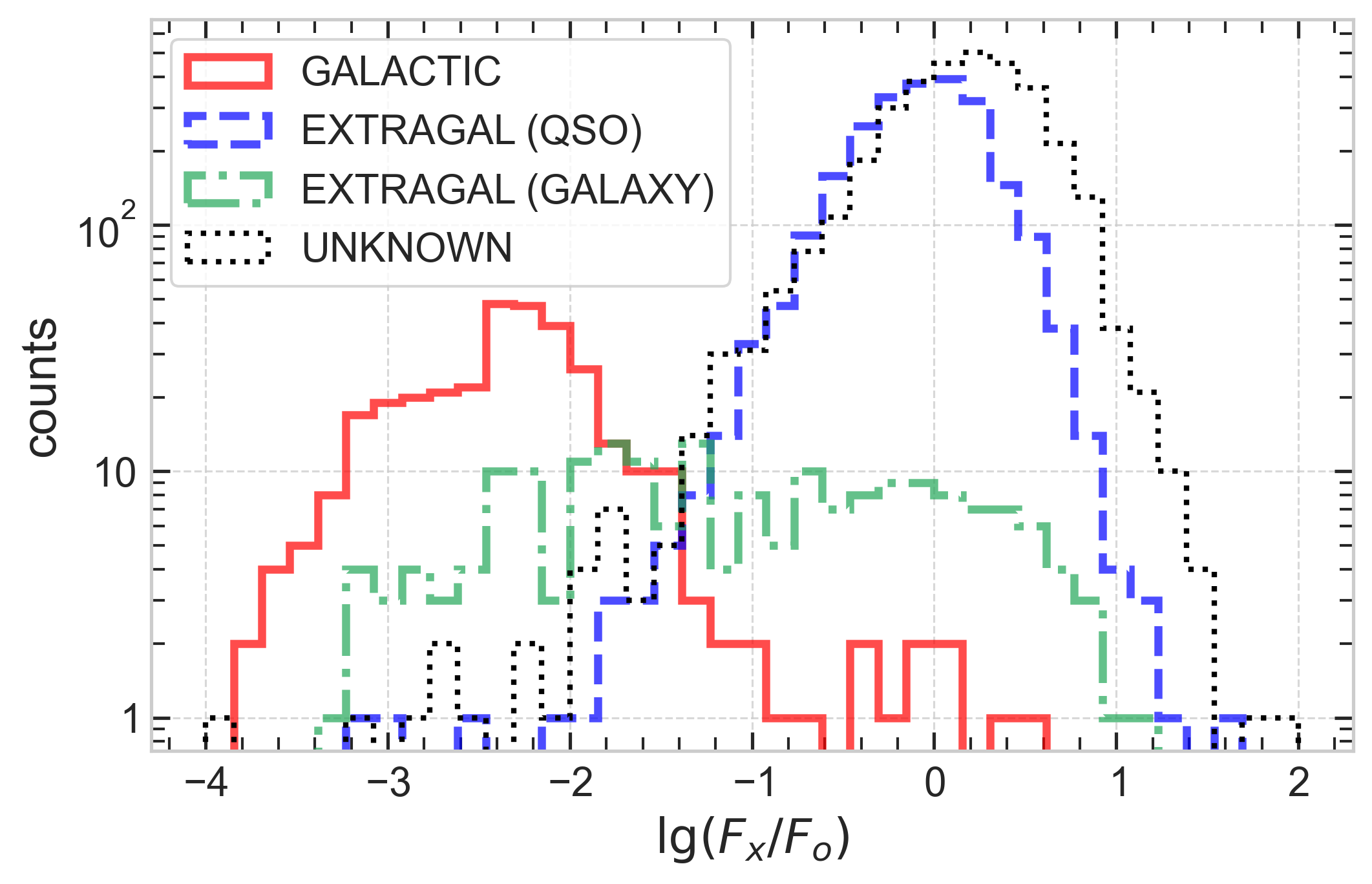}
    \caption{The $\lg(F_{\rm x} / F_{\rm o})$ distributions for the Galactic (red solid line) and extragalactic (the green dash-dotted and blue dashed lines for the galaxies and quasars, respectively) sources. The distribution of the sources without a class is indicated by the dotted line. The classification of sources is based on the Gaia catalog, the spectroscopic catalogs, and the Simbad database, as described in the text. It can be seen that the ratio of the X-ray and optical fluxes separates satisfactorily the Galactic sources and quasars. At the same time, additional information is required to separate the Galactic sources and galaxies.
    }
    \label{fig:FxFo_valid_hist}
    \end{figure*}
    \item A small fraction of the extragalactic objects (mostly galaxies\footnote{Note that the eROSITA X-ray sources classified as “galaxy”, as a rule, have an active nucleus of low or moderate luminosity, so that the emission from the stellar population plays a prominent role in the optical band. Their emission in the X-ray band is determined by accretion onto a supermassive black hole.}) penetrate into the left part of the diagram near $r - z \sim 0.5$. In Fig. \ref{fig:FxFo_rz} this region is marked by the dashed line in the form of a triangle. However, it can be seen that some of the extragalactic objects can also be located below the designated region. It can also be seen that the triangular region is uniformly filled with stars: their density in it does not decrease compared to other parts of the left half of the diagram.
    
    It is important to note that 81\% of the unclassified objects in the left part of the diagram are optically extended ones (gray circles with black boundary), suggesting their extragalactic nature. This can be explained by the fact that the extragalactic objects penetrating into the left part of the diagram are, as a rule, relatively nearby galaxies that are comparatively easily resolved by modern optical telescopes. This is illustrated by Fig. \ref{fig:z_extra_hist}, which shows the redshift distribution of the extragalactic objects from different parts of the $\lg(F_{\rm x} / F_{\rm o})$–color diagram.

    \begin{figure*}
    \centering
    \includegraphics[scale=0.6]{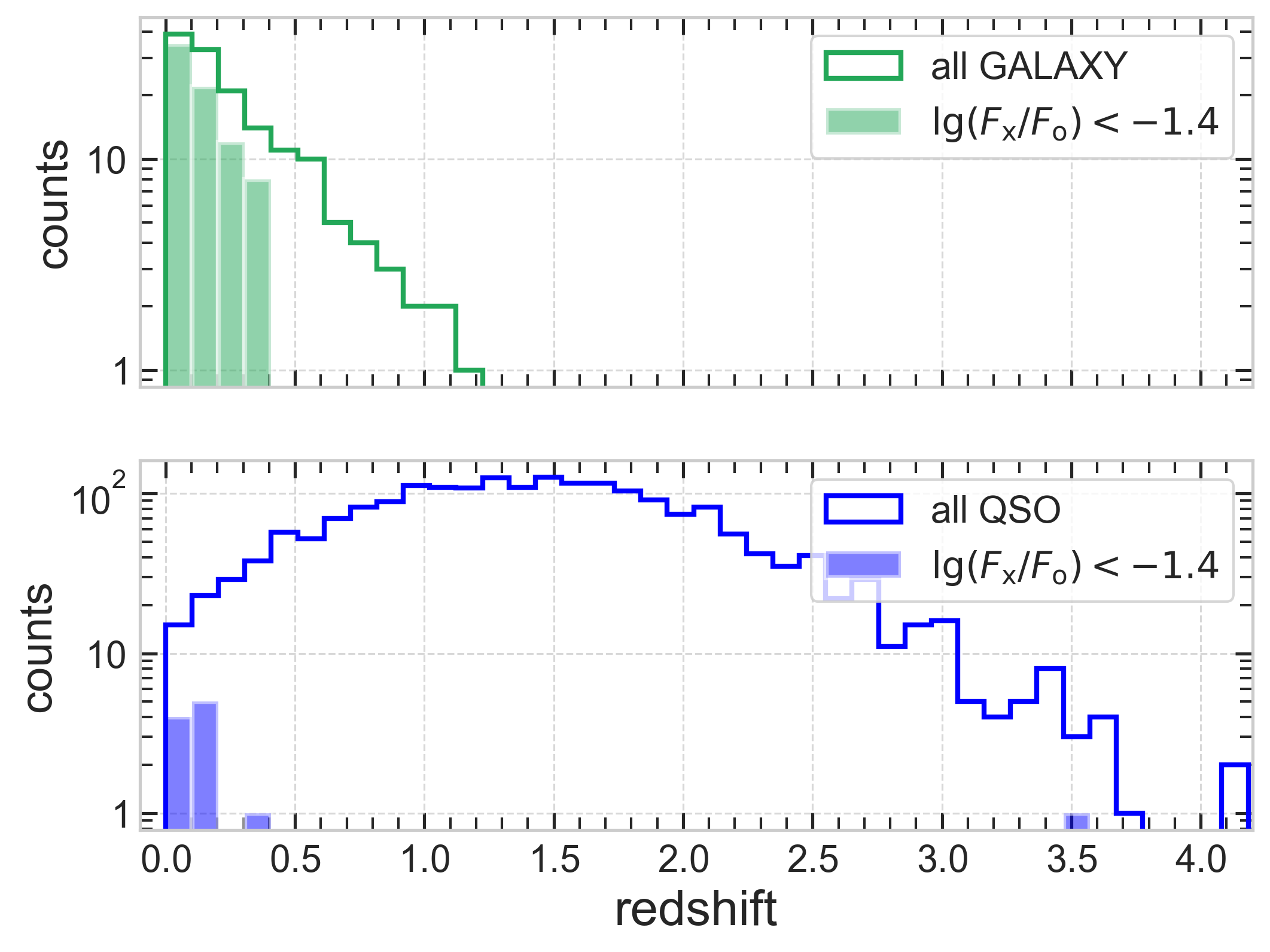}
    \caption{Redshift distribution of the extragalactic objects from different parts of the diagram in Fig. \ref{fig:FxFo_rz}. The distributions for the galaxies and quasars are presented on the upper and lower panels, respectively. On both panels the lines and the shaded columns designate the distributions of all sources of the specified type and the distributions of the corresponding types of objects from the left part of the diagram in Fig. \ref{fig:FxFo_rz}, respectively ($\lg(F_{\rm x} / F_{\rm o}) < -1.4$). It can be seen that low-redshift objects fall into the left part of the diagram in Fig. \ref{fig:FxFo_rz}.}
    \label{fig:z_extra_hist}
    \end{figure*}

    \item A small number (24) of Galactic objects are located in the right part of the $\lg(F_{\rm x} / F_{\rm o})$--color diagram. These Galactic objects have high $\lg(F_{\rm x} / F_{\rm o})$ atypical for stars and, apparently, are cataclysmic variables and/or active binaries etc. Some of them located near the boundary can be M-type stars. Interestingly, half (12) of the objects with high $\lg(F_{\rm x} / F_{\rm o})$ were identified based on the spectroscopic data rather than the Gaia measurements. One of the next papers in the series of publications on the Lockman Hole will be devoted to a detailed study and classification of these objects.
\end{enumerate}

\subsection{Classification Algorithm}\label{sect-method-algorithm}

Using our analysis of the diagram in Fig. \ref{fig:FxFo_rz}, we formulated an algorithm for the classification of point-like X-ray sources from the Lockman Hole catalog. This algorithm does not claim to be universal and was tailored to the specific parameters of the deep Lockman Hole survey with the eROSITA telescope. Second, our goal at this stage is the binary classification of the X-ray sources in the Lockman Hole into Galactic and extragalactic sources. The classification algorithm based on the ideas presented here will be generalized in our succeeding papers.

The classification scheme includes the following four steps:

\begin{enumerate}
  \item The Galactic objects were classified based on the Gaia data (338) or the classification of the spectroscopic catalogs (13).
  \item The extragalactic sources were classified based on the classification of the spectroscopic catalogs and Simbad (2339 quasars, 194 galaxies).
\end{enumerate}

The sources to which we failed to assign a class based on the Gaia data or the spectroscopic catalogs were processed as follows:

\begin{enumerate}[resume]
    \item \label{extended_unknown} All of the optically extended sources (in the DESI LIS catalog $\text{type} \neq \text{PSF}$) were classified as extragalactic (1804 sources).
    \item \label{fxfo} The point sources were classified in accordance with $\lg(F_{\rm x} / F_{\rm o})$. The sources were classified as extragalactic at $\lg(F_{\rm x} / F_{\rm o}) > -1.4$ (1592 sources) and as Galactic at $\lg(F_{\rm x} / F_{\rm o}) < -1.4$ (6 sources).
\end{enumerate}

Step \ref{fxfo} of this scheme is based on the fact that the extragalactic sources account for more than 99.9\% of all the sources with a known class rightward of the vertical line in Fig. \ref{fig:FxFo_rz} (i.e., with $\lg(F_{\rm x} / F_{\rm o}) > -1.4$), while 99.7\% of all the point sources leftward of this line are Galactic. Note that this classification leads to the “loss” of Galactic objects with high $\lg(F_{\rm x} / F_{\rm o})$. However, within the binary classification, which is our goal, this loss is insignificant, as follows from the above numbers. This is discussed in more detail in the next section. The number of extended and point sources for each class is given in Table \ref{table:extended_stat}.

\begin{table}[t]
\centering
\resizebox{0.85 \columnwidth}{!}{
\begin{tabular}{lrr}
\toprule
{} & \multicolumn{2}{l}{} \\
Source type & Point &  Extended \\
\hline
\midrule
Galactic            &   333 &     5 \\
Galactic (non-Gaia) &     12 &     1 \\
Galaxy              &    28 &   166 \\
Quasar                 &  1886 &   453 \\
Unknown             &  1626 &  1836 \\
\bottomrule
\end{tabular}
}
\caption{The number of optically extended and point sources separated into classes. Almost all of the Galactic objects are point ones. The optically extended Galactic objects are discussed in \ref{result-peculiar}.}
\label{table:extended_stat}
\end{table}

\begin{figure*}
\centering
\includegraphics[scale=0.6]{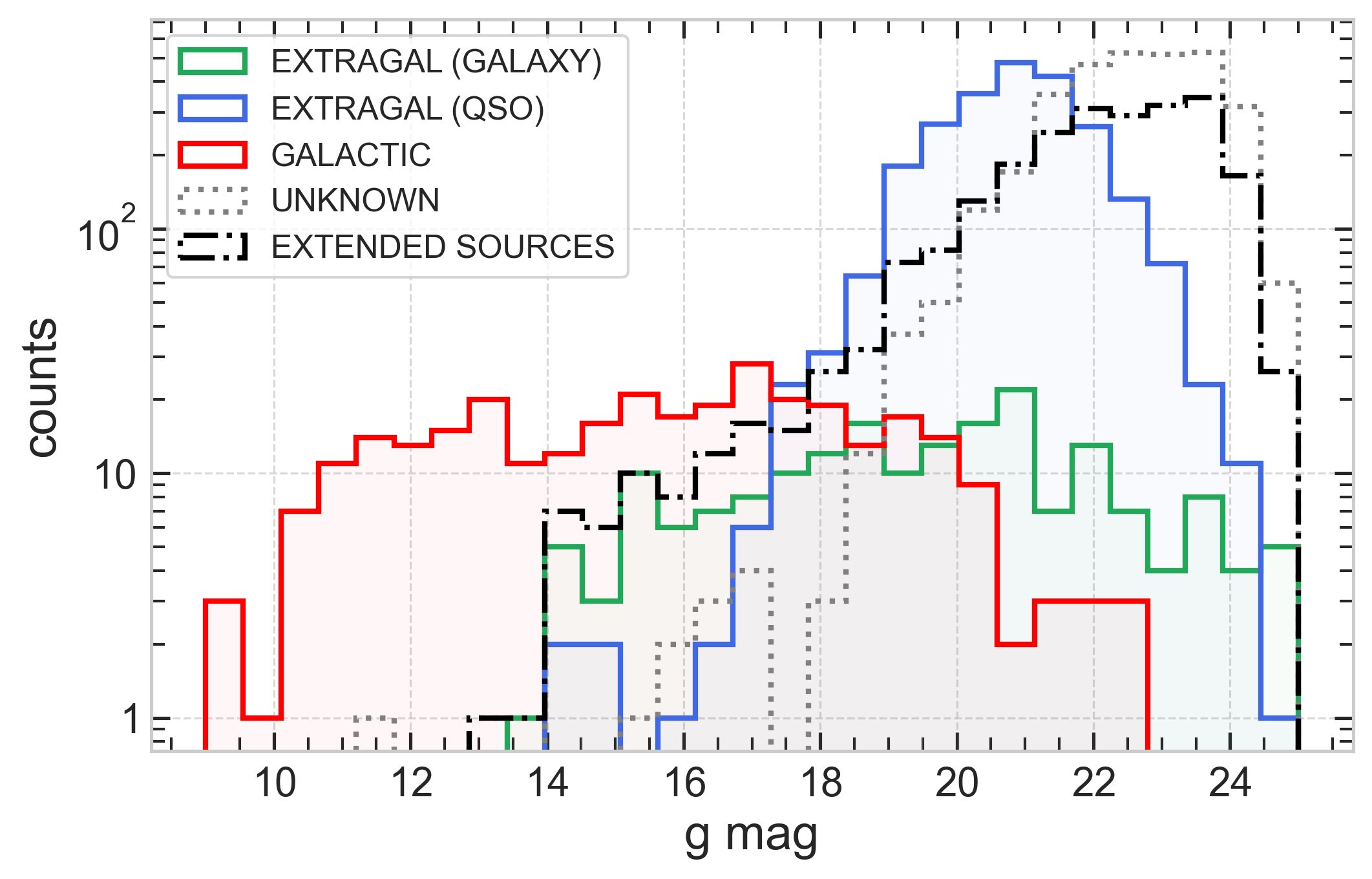}
\caption{
The $g$ magnitude distributions for the Galactic and extragalactic sources described in Subsection \ref{sect-method-algorithm}. The Galactic objects (red solid line) are, on average, brighter than the extragalactic ones (the blue and green lines for the quasars and galaxies, respectively). The black dash–dotted line indicates the distribution for the optically extended sources. The unclassified objects (gray dotted line) are, on average, less bright than those to which we managed to assign a class based on the Gaia catalog, the spectroscopic catalogs, and Simbad.
}
\label{fig:gmag_valid_hist}
\end{figure*}

Figure \ref{fig:gmag_valid_hist} presents the $g$ magnitude distributions for the sources mentioned in different steps of the classification algorithm. The unclassified objects (gray dotted line) are, on average, less bright in the optical range than those to which we managed to assign a class based on the Gaia catalog and the spectroscopic catalogs (for more details, see Subsection \ref{sect-metrics}).

Of the \ntotal sources from the catalog, we considered \noverpany sources with a counterpart ($p_{\rm any} > 0.12$). For \nwithclasses of them we managed to assign a class based on the extraneous catalogs or the Simbad database (see Subsection \ref{sect-method-gaiaetc}). Of the remaining \nunknown objects, \nunknownnozmag have an optical $z$ flux in the DESI LIS catalog with S/N $<$ 3. These objects were not classified, since $F_{\rm x} / F_{\rm o}$ for them were deemed unknown. We classified 3402 sources according to the rules described above. Table \ref{table:classification_stat} gives the final number of extragalactic and Galactic sources.

\begin{table}[t]
\centering
\resizebox{0.9 \columnwidth}{!}{
\begin{tabular}{ccc}
\begin{tabular}[c]{@{}c@{}}Classification\\ method\end{tabular} & Extragalactic & Galactic \\
\hline
\begin{tabular}[c]{@{}c@{}}Based on catalogs\\ (2884)\end{tabular}  & 2533             & 351           \\
\begin{tabular}[c]{@{}c@{}}According to rules\\ (3402)\end{tabular}   & 3396             & 6            \\
\begin{tabular}[c]{@{}c@{}}Total\\ (6286)\end{tabular}   & 5929             & 357            
\end{tabular}
}
\caption{The number of extragalactic and Galactic sources among the objects to which we assigned a class based
on the extraneous catalogs or the Simbad database (see Subsection \ref{sect-method-gaiaetc}) and those classified according to the rules described in Subsection \ref{sect-method-algorithm}}
\label{table:classification_stat}
\end{table}

\subsection{Classification Quality Metrics} \label{sect-metrics}

To estimate the classification quality, we used the sample of \nvalidation objects to which we managed to assign a class based on the Gaia catalog, the spectroscopic catalogs, and the Simbad database and the optical $z$ flux from which was determined reliably (hereafter the validation sample).

However, it is worth noting that the validation sample constructed in this way is not quite representative, since it is subject to selection effects. This is primarily because for the spectroscopic SDSS programs the objects satisfying certain criteria in accordance with the peculiar goals of these observing programs were chosen. In addition, reliable (and, on average, brighter) targets are usually chosen for spectroscopic observations. For this reason, the validation sample contains, on average, brighter sources (see Fig. \ref{fig:gmag_valid_hist}). Larger validation samples are required to achieve the necessary accuracy of determining the quality metrics, whereas there exist a limited number of sources of spectroscopic information for a large number of objects. Therefore, the construction of a representative validation sample is a general difficulty that the creators of the object classification methods run into; this problem is beyond the scope of our paper.

Our validation sample allows the quality of the binary classification into Galactic and extragalactic objects to be estimated. For this purpose, we applied steps \ref{extended_unknown} and \ref{fxfo} of the classification algorithm to the sources from this sample in such a way as if the classes of these sources are unknown to us. The results of our comparison of the predicted and real classes are discussed below.

Errors of two types can be made during the classification: we can erroneously call an extragalactic source a Galactic one and vice versa. Let us designate the extragalactic objects as a positive class. The Galactic object erroneously classified as extragalactic will then give a false positive (FP) result. The Galactic object classified as extragalactic will give a false negative (FN) result. The recall for the classification of extragalactic sources is calculated from the formula TP/(TP + FN), while the precision is calculated as TP/(TP + FP), where TP (true positive) is the number of correctly classified extragalactic sources. The metrics for the Galactic sources as a
positive class are calculated in a similar way.

On the validation sample we managed to correctly classify 2529 extragalactic objects and 306 Galactic ones. At the same time, 28 Galactic objects were erroneously classified as extragalactic and one extragalactic object was classified as Galactic. Thus, the recall and the precision are

\begin{itemize}
    \item $99.96 \pm0.04 \%$ and $98.90 \pm0.2\%$, respectively, for the selection of extragalactic objects,
    \item $91.62 \pm0.5\%$ and $99.67 \pm0.1\%$, respectively, for the selection of Galactic objects.
\end{itemize}

The statistical uncertainties given above were calculated under the assumption of a binomial distribution.

We also repeated the above calculations by dividing the sample into two halves and using one of them to choose the parameters of the classification algorithm and the other one to determine the recall and precision metrics and obtained results similar to those described above.

In accordance with the discussion in the previous subsection (see the discussion after step \ref{fxfo} of the classification scheme), 28 of the 29 incorrectly classified objects are accounted for by the cases where a Galactic object is erroneously classified as extragalactic. Among them there are almost all (12 of 13) of the
objects from the validation sample that were found to belong to the Galactic ones \textit{not} based on the Gaia data (see Subsection \ref{result-peculiar} and Table \ref{table:non_gaia_stars}). They all have atypically high values of $F_{\rm x} / F_{\rm o}$.

\section{DISCUSSION}\label{sect-results}
\subsection{Curves of Cumulative Number Counts (logN--logS) for the Galactic and Extragalactic Sources}\label{sect-logNlogS}

\begin{figure}
\centering
\includegraphics[width=\columnwidth]{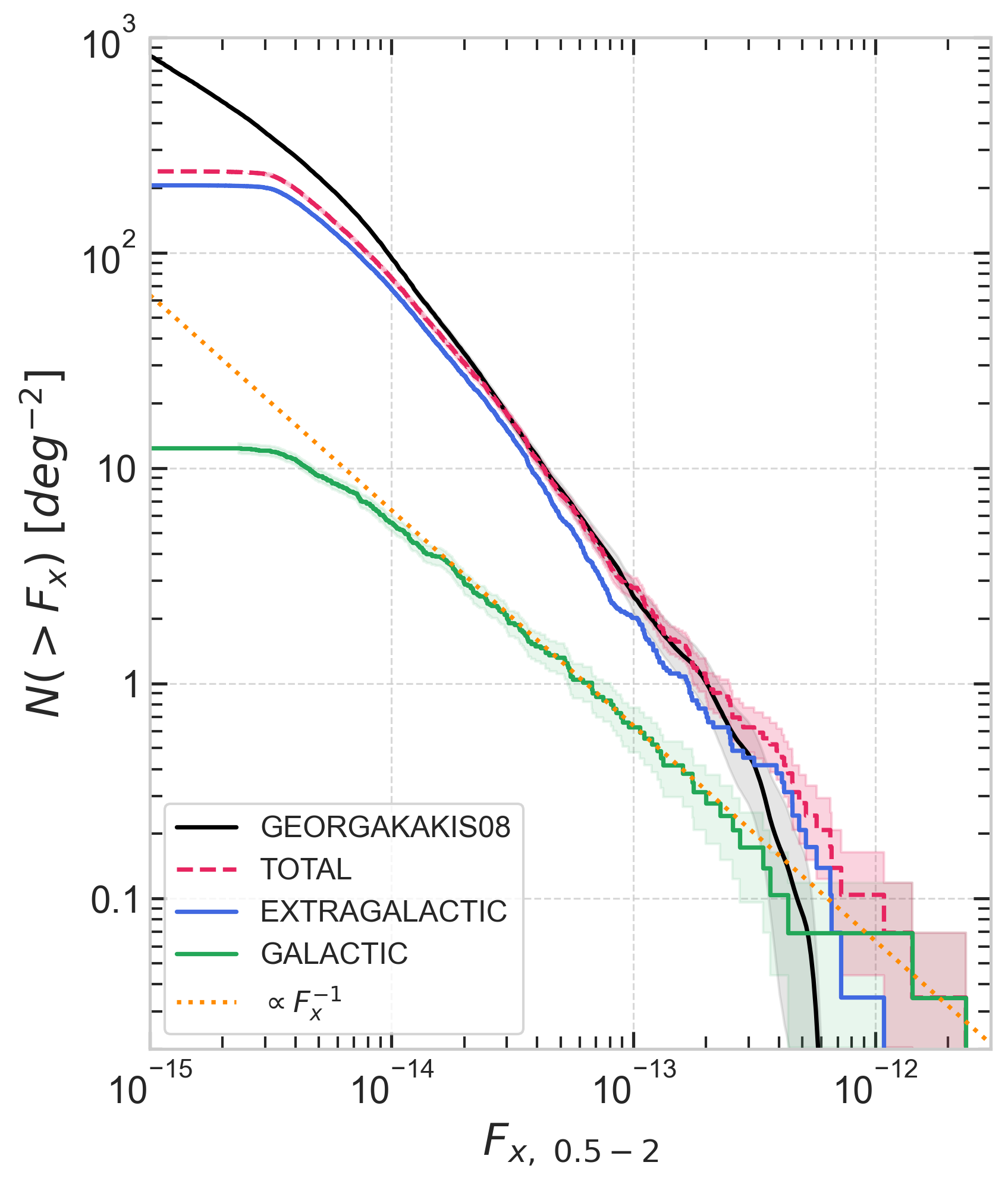}
\caption{
The curves of cumulative source number counts log$N$--log$S$ in the Lockman Hole. The curve was normalized to square degree. The red dashed line indicates the curve for all point X-ray sources from the catalog by Gilfanov et al. (2023, in preparation). The blue and green solid lines indicate the curves of cumulative number counts for the sources classified as extragalactic and Galactic, respectively. The dotted line indicates the dependence $\propto F_X^{-1}$, normalized to the number of Galactic sources at fluxes $\gtrsim 2\times 10^{-14}$ \ergcms. The semitransparent regions around the curves indicate the statistical uncertainties calculated under the assumption of a Poisson distribution. The black line indicates the curve of cumulative source number counts from \citet{Georgarakis08}.
}
\label{fig:logNlogS}
\end{figure}

Figure \ref{fig:logNlogS} presents the curves of cumulative number counts\footnote{Recall that the curve of cumulative source number counts (the so-called log$N$--log$S$ curve) shows the flux dependence of the number of sources (as a rule, normalized to unit solid angle) with a flux higher than the specified one.} for the sources classified as Galactic and extragalactic and the full curve of cumulative number counts for all sources. For comparison, the figure presents the curve of cumulative number counts from the Chandra data in \citet{Georgarakis08}. For the eROSITA sources the counts were converted to the 0.5–2.0 keV flux corrected for interstellar absorption by assuming a median value of $NH=7\times 10^{19}$ cm$^{-2}$ and a slope of the power-law spectrum $\Gamma=1.9$.

The incompleteness of the curves of cumulative number counts presented in Fig. \ref{fig:logNlogS} was not corrected in this paper. It follows from the curve of the effective Lockman Hole survey area given in \citep{gilfanov2022} that the incompleteness effects begin to play a role at a $\sim (2-3)\times 10^{-14}$ \ergcms. In Fig. \ref{fig:logNlogS} this is reflected in the flattening of the curves of cumulative number counts at lower fluxes.

Extragalactic sources (AGNs and quasars) account for most of the sources detected by eROSITA, while the fraction of Galactic sources (stars) at fluxes $\sim10^{-14}$ \ergcms is $\sim10\%$. On the whole, this value is typical for fields at high Galactic latitudes, but it can change with location in the sky. A difference in the slopes of the curves of cumulative number counts for these two populations is also clearly seen. The curve of cumulative number counts for the extragalactic sources follows the law $\propto F_X^{-3/2}$, as would be expected in this range of fluxes, whereas the curve of cumulative number counts for the Galactic sources is flatter and follows the dependence $\propto F_X^{-1}$. Note that this slope coincides remarkably well with the value expected for the population of sources in the disk, but this analogy seems to be inapplicable here.

\citet{Georgarakis08} did not separate the sources into extragalactic and Galactic ones. Accordingly, their curve of cumulative number counts agrees well with the full eROSITA curve of cumulative number counts. For example, the number density of sources at fluxes $F_{\rm x}>2\times 10^{-14}$ is $33.2\pm1.1$ and $33.6\pm1.4$ sources/deg$^2$ from the eROSITA and Chandra data, respectively.

\subsection{Peculiar Sources}\label{result-peculiar}

Here we provide the lists of objects that for some reasons are unusual and were revealed during the classification of X-ray sources in the Lockman Hole.

Table \ref{table:star_outliers} lists the Gaia Galactic sources selected in Subsection \ref{sect-method-gaiaetc} and lying in Fig. \ref{fig:FxFo_rz} in the region of extragalactic objects ($\lg(F_{\rm x} / F_{\rm o}) > -1.4$). These Galactic sources have an atypically large ratio of the X-ray flux to the optical one and can be candidates for cataclysmic variables. The source SRGe J103224.9+572814 is worth noting separately: according to the Gaia catalog, it has a significant proper motion ($12.5 \pm 0.5$ mas yr$^{-1}$), while being a quasar ($z = 2.43 \pm 0.00014$), according to the spectroscopic catalogs. For more details on such objects, see, e.g., \cite{Souchay2022} and \cite{hamitov2022}.

Table \ref{table:non_gaia_stars} lists the objects that were classified as Galactic based on the spectroscopic catalogs or the Simbad database (without using Gaia data). Most of them, just as the objects from Table \ref{table:star_outliers}, have $\lg(F_{\rm x} / F_{\rm o})$ typical for extragalactic objects.

The optical extent of some of the Gaia Galactic objects (marked by the red circles with the black boundary in Fig. \ref{fig:FxFo_rz}, their list is given in Table \ref{table:extended_stars}) can be associated with their relatively large proper motions and, as a consequence, the source position displacement between the optical images taken at different times. In turn, the proper motions can be caused by the galaxy photocenter displacement due to, for example, the motion of the jets or microlensing. The errors of the Gaia astrometric solutions cannot
be ruled out completely either. The recent paper by \citet{hamitov2022} is devoted to searching for and studying such objects in the eROSITA X-ray catalog.

\begin{table*}[t]
\resizebox{\textwidth}{!}{
\begin{tabular}{lllllrr}
\toprule
{} &            Source & SDSS class & MILQ class &                    Simbad class & Color & lg(Fx/Fo)\\
\midrule
\hline
1  &  SRGe J103906.0+553044 &         -  &         -  &                            Star &     -3.62 & -1.37 \\
2  &  SRGe J105633.1+583529 &         -  &         -  &                            Star &     -1.82 & -1.30 \\
3  &  SRGe J104522.1+555738 &         -  &        QSO &                    Radio Source &     -0.56 & -1.27 \\
4  &  SRGe J103618.3+581246 &         -  &         -  &                              -  &     -0.70 & -1.21 \\
5  &  SRGe J104007.7+595700 &         -  &         -  &  Long-Period Variable Candidate &     -2.56 & -1.15 \\
6  &  SRGe J103453.6+553633 &         -  &         -  &                              -  &      1.81 & -1.14 \\
7  &  SRGe J105310.4+575437 &         -  &         -  &                              -  &      1.76 & -1.01 \\
8  &  SRGe J103224.9+572814 &        QSO &        QSO &                          Quasar &      1.11 & -0.93 \\
9  &  SRGe J105848.1+592918 &         -  &         -  &                      HII Region &     -0.60 & -0.82 \\
10 &  SRGe J102411.6+561606 &         -  &         -  &                              -  &     -0.36 & -0.64 \\
11 &  SRGe J110137.5+572926 &         -  &         -  &                            Star &     -2.87 &  0.03 \\
12 &  SRGe J104325.5+563300 &         -  &         -  &              Cataclysmic Binary &     -0.32 &  0.13 \\
\bottomrule
\end{tabular}
}
\caption{
The Gaia Galactic sources (designated by the red circles in Fig. \ref{fig:FxFo_rz}) with $\lg(F_{\rm x} / F_{\rm o}) > -1.4$. The colors below --1 may be distorted and are discussed in Subsection \ref{sect:fxfo_color}.
}
\label{table:star_outliers}
\end{table*}

\begin{table*}[t]
\resizebox{\textwidth}{!}{
\begin{tabular}{lllllllr}
\toprule
{} &            Source & SDSS class & SDSS subclass & MILQ class & Simbad class & Color & lg(Fx/Fo) \\
\midrule
\hline
1  &  SRGe J103049.0+563614 &         -  &            -  &         -  &         Star &     -0.02 & -0.27 \\
2  &  SRGe J103606.5+573624 &       STAR &            K5 &       STAR &           -  &      0.34 & -0.08 \\
3  &  SRGe J104809.2+581036 &         -  &            -  &         -  &         Star &     -0.41 & -3.59 \\
4  &  SRGe J105106.4+552342 &       STAR &       O9.5Iae &       STAR &           -  &      0.08 &  0.31 \\
5  &  SRGe J105237.2+600836 &         -  &            -  &         -  &         Star &      0.24 & -0.03 \\
6  &  SRGe J103256.5+574818 &         -  &            -  &         -  &         Star &      0.78 &  0.53 \\
7  &  SRGe J103511.4+600547 &         -  &            -  &       STAR &           -  &      0.16 &  0.05 \\
8  &  SRGe J105118.5+551934 &       STAR &       O9.5Iae &       STAR &  Blue Object &     -0.15 & -0.12 \\
9  &  SRGe J104636.2+555934 &         -  &            -  &         -  &         Star &     -1.43 & -0.01 \\
10 &  SRGe J103426.1+575524 &         -  &            -  &         -  &         Star &        -  &  0.46 \\
11 &  SRGe J105432.9+590945 &         -  &            -  &         -  &         Star &       0.7 & -0.29 \\
12 &  SRGe J103744.3+571156 &       STAR &           O8e &        QSO &       BL Lac &      0.49 &  0.43 \\
13 &  SRGe J104429.8+595354 &         -  &            -  &         -  &         Star &      0.14 & -0.38 \\
\bottomrule
\end{tabular}
}
\caption{
The objects that were classified as Galactic based on information from the spectroscopic catalogs or the Simbad database (not based on Gaia data). Marked by the red circumferences in Fig. \ref{fig:FxFo_rz}. The colors below --1 may be distorted and are discussed in Subsection \ref{sect:fxfo_color}.
}
\label{table:non_gaia_stars}
\end{table*}

\begin{table*}[t]
\centering
\resizebox{13cm}{!}{
\begin{tabular}{llrrrr}
\toprule
{} &            Source & GAIA PM (mas/yr) & GAIA PM SNR & lg(Fx/Fo) \\
\midrule
\hline
1 &  SRGe J105146.0+554552 &           144.64 &      204.3 &      1.91 \\
2 &  SRGe J104425.9+565729 &            24.21 &       69.3 &      1.46 \\
3 &  SRGe J105603.0+595317 &             5.58 &       14.4 &     -0.24 \\
4 &  SRGe J105130.7+573439 &            63.28 &      310.6 &      2.00 \\
5 &  SRGe J102411.6+561606 &             9.38 &      602.8 &     -0.36 \\
\bottomrule
\end{tabular}
}
\caption{
The Gaia Galactic objects that are extended objects according to the DESI LIS catalog. 
Marked by the red circumferences with the black boundary in Fig. \ref{fig:FxFo_rz}.
}
\label{table:extended_stars}
\end{table*}


\section{CONCLUSIONS}\label{sect-ending}

We classified the X-ray sources in the \srge catalog produced based on the Lockman Hole survey \citep{gilfanov2022}. For this purpose, we used the results of the crossmatch of X-ray sources with objects from the optical DESI LIS catalog made in \cite{bykov}. Of the \ntotal X-ray sources, we managed to assign a class to \nwithclasses sources after their cross-match with the Gaia DR3, SDSS DR16, MILLIQUAS catalogs, and the Simbad database. For the remaining sources we performed the classification based on information about the object optical extent and the ratio $F_{\rm x}/F_o$ of the X-ray and optical ($z$) fluxes. As a
result, we classified \ngal and \nextragal sources as Galactic and extragalactic, respectively. We estimated the selection precision and recall for the extragalactic (99.9 and 98.9\%, respectively) and Galactic (91.6 and 99.7\%) sources. No classification was made for the \nunknownnozmag sources for which the DESI LIS photometry had an insufficient S/N. Using the cross-match results from \cite{bykov}, we classified \nhostless sources as hostless, i.e., having no optical counterparts in the DESI LIS survey. The results of this paper will be used in our subsequent papers to construct the luminosity function of AGNs and quasars in the Lockman Hole field. The code that accompanies this paper is available at \url{https://github.com/mbelveder/ero-lh-class.git}.

\newpage

\section*{ACKNOWLEDGMENTS}\label{sect-thanks}

This study is based on observations with the eROSITA telescope onboard the SRG observatory. The SRG observatory was built by Roskosmos in the interests of the Russian Academy of Sciences represented by the Space Research Institute (IKI) within the framework of the Russian Federal Space Program, with the participation of the Deutsches Zentrum fur Luft- und Raumfahrt (DLR). The SRG/eROSITA X-ray telescope was built by a consortium of German institutes led by MPE, and supported by DLR. The SRG spacecraft was designed, built, launched and is operated by the Lavochkin Association and its subcontractors. The science data are downlinked via the Deep Space Network Antennae in Bear Lakes, Ussurijsk, and Baykonur, funded by Roskosmos. The eROSITA data used in this paper were processed with the eSASS software developed by the German eROSITA consortium and the proprietary data reduction and analysis software developed by the Russian eROSITA consortium.

The following software was used: NumPy \citep{Harris2020}, Matplotlib \citep{Hunter2007},  SciPy \citep{2020SciPy-NMeth}, Pandas\citep{reback2020pandas},  AstroPy \citep{astropy:2018}. We used the SIMBAD data base maintained by the Strasbourg Astronomical Data Center (CDS\footnote{\url{https://simbad.u-strasbg.fr/simbad}}, Strasbourg, France).


\vfill
 
\bibliographystyle{astl}
\bibliography{refs_rus.bib}

\end{document}